# Application of Helmholtz-Hodge decomposition and conditioned structure functions to exploring influence of premixed combustion on turbulence upstream of the flame

Vladimir A. Sabelnikov[1,2], Andrei N. Lipatnikov[3], Nikolay Nikitin[4], Shinnosuke Nishiki[5], Tatsuya Hasegawa[6]

[1]*ONERA – The French Aerospace Laboratory, F-91761 Palaiseau, France*
[2]*Central Aerohydrodynamic Institute (TsAGI), 140180 Zhukovsky, Moscow Region, Russian Federation*
[3]*Department of Mechanics and Maritime Sciences, Chalmers University of Technology, Gothenburg, 41296 Sweden*
[4]*Moscow State University, Moscow Region, Russian Federation*
[5]*Department of Information and Electronic Engineering, Teikyo University, Utsunomiya 320-8551, Japan,*
[6]*Institute of Materials and Systems for Sustainability, Nagoya University, Nagoya 464-8603, Japan*

**Corresponding author:** Prof. Vladimir Sabelnikov, ONERA – The French Aerospace Laboratory, F-91761 Palaiseau, France, vladimir.sabelnikov@onera.fr

**Abstract**

In order to explore the influence of combustion-induced thermal expansion on turbulence, a new research method is introduced. The method consists in jointly applying Helmholtz-Hodge decomposition and conditioned structure functions to analyzing turbulent velocity fields. Opportunities offered by the method are demonstrated by using it to process Direct Numerical Simulation data obtained earlier from two statistically 1D, planar, fully-developed, weakly turbulent, single-step-chemistry, premixed flames characterized by two significantly different (7.52 and 2.50) density ratios, with all other things being approximately equal. To emphasize the influence of combustion-induced thermal expansion on turbulent flow of unburned mixture upstream of a premixed flame, the focus of analysis is placed on structure functions conditioned to the unburned mixture in both points. Two decomposition techniques, i.e. (i) a widely used orthogonal Helmholtz-Hodge decomposition and (ii) a recently introduced natural Helmholtz-Hodge decomposition, are probed, with results obtained using them being similar in the largest part of the computational domain with the exception of narrow zones near the inlet and outlet boundaries. Computed results indicate that combustion-induced thermal expansion can significantly change turbulent flow of unburned mixture upstream of a premixed flame by generating anisotropic potential velocity fluctuations whose spatial structure differ substantially from spatial structure of the incoming turbulence. The magnitude of such potential velocity fluctuations is greater than the magnitude of the solenoidal velocity fluctuations in the largest part of the mean flame brush in the case of the high density ratio. In the case of the low density ratio, the latter magnitude is larger everywhere, but the two magnitudes are comparable in the middle of the mean flame brush. Contrary to the potential velocity fluctuations, the influence of the thermal expansion on the solenoidal velocity field in the unburned mixture is of minor importance under conditions of the present study.

**1. Introduction**

Since an increase in burning rate due to flame-generated turbulence was hypothesized by Karlovitz et al. [1] and Scurlock and Grover [2], the influence of combustion-induced density variations on turbulent flow within a premixed flame brush was studied in a number of papers reviewed elsewhere [3-6], see also recent Refs. [7-11]. The majority of such studies addressed the first and second moments of velocity field within mean flame brush, with the moments being averaged in the conventional way by allowing for contributions from unburned reactants, intermediate mixture states, and combustion products simultaneously. However, as far as eventual increase in burning rate due to flame-generated turbulence is concerned, combustion-induced perturbations of the flow upstream of the flame appear to be of the most importance. Indeed, since a flame propagates into unburned gas, velocity perturbations upstream of the flame are required to affect the flame propagation.

To distinguish the influence of thermal expansion on turbulence in unburned gas, reacting mixture, or products, conditioned flow characteristics are commonly used, as reviewed elsewhere [4-6]. Recently, Structure Functions (SFs) of the velocity field, conditioned to different mixture states, were introduced for that purpose in Refs. [12,13] and [14] using two different frameworks. Results obtained from weakly turbulent flames [12,13] indicated that combustion could significantly affect turbulent flow of unburned reactants not only within, but also upstream of the mean flame brush. Such effects were attributed to potential flow fluctuations caused by combustion-induced pressure perturbations upstream of the mean flame brush. The goal of the present work is to examine this hypothesis by decomposing the flow field into potential (irrotational) and solenoidal (rotational) subfields and separately exploring characteristics of each subfield.

Direct Numerical Simulation (DNS) data analyzed in the following are summarized in Section 2. Methods applied to process these data are discussed in Section 3. Results are presented in Section 4, followed by conclusions.

**2. DNS Attributes**

In the present study, DNS data computed by Nishiki et al. [15,16] almost 20 years ago and used in the aforementioned papers [12,13] are further analyzed. The choice of this database, which appears to be outdated when compared to recent DNS data [17-22] generated invoking complex combustion chemistry at a high ratio of the rms turbulent velocity $u'$ to the laminar flame speed $S_L$, requires comments. The point is that the focus of the present study is placed on the influence of combustion-induced thermal expansion on the turbulent flow upstream of a flame. Accordingly, modeling of intermediate states of the mixture (e.g., complex combustion chemistry) appears to be of secondary importance when compared to two other major requirements. First, to clearly reveal thermal expansion effects, data obtained at different density ratios $\sigma = \rho_u/\rho_b$ are very useful and the DNS data analyzed here were obtained at $\sigma = 2.5$ or 7.53, with all other things being roughly equal. Second, a new research method (e.g. conditioned SFs extracted from potential and solenoidal velocity subfields) should initially be probed under conditions associated with the strongest manifestation of effects (e.g. combustion-induced perturbations of the potential velocity subfield upstream of a flame) the method aims at. To make such effects as strong as possible, the flamelet regime of premixed turbulent burning [23], associated with a low ratio of $u'/S_L$, is of the most interest. The DNS by Nishiki et al. [15,16] did deal with the flamelet regime, as discussed in detail elsewhere [24], whereas the majority of recent DNS studies explored other combustion regimes.

Because the DNS data analyzed here were extensively discussed by various research groups [12,13,15,16,24-43], we will restrict ourselves to a brief summary of those compressible simulations. They dealt with statistically 1D, planar, adiabatic flames modeled by unsteady 3D continuity, Navier-Stokes, and energy equations, supplemented with a transport equation for the mass fraction $Y$ of a deficient reactant and the ideal gas state equation. The Lewis and Prandtl numbers were equal to 1.0 and 0.7, respectively, and combustion chemistry was reduced to a single reaction. Accordingly, the combustion progress variable $c(\mathbf{x}, t) = 1 - Y(\mathbf{x}, t)/Y_u = (T(\mathbf{x}, t) - T_u)/(T_b - T_u)$.

The computational domain was a rectangular box $\Lambda_x \times \Lambda_y \times \Lambda_z$ with $\Lambda_x = 8$ mm, $\Lambda_y = \Lambda_z = 4$ mm, and was resolved using a uniform rectangular ($2\Delta x = \Delta y = \Delta z$) mesh of $512 \times 128 \times 128$ points. The flow was periodic in $y$ and $z$ directions. Homogeneous isotropic turbulence ($u' = 0.53$ m/s, an integral length scale $L = 3.5$ mm, the turbulent Reynolds number $Re_t = 96$ [15,16]) was generated in a separate box and was injected into the computational domain through the left boundary $x = 0$. In the domain, the turbulence decayed along the direction $x$ of the mean flow.

At $t = 0$, a planar laminar flame was embedded into statistically the same turbulence assigned for the velocity field in the entire computational domain. Subsequently, the inflow velocity was increased twice, i.e., $U(0 \le t < t_I) = S_L < U(t_I \le t < t_{II}) < U(t_{II} \le t) = S_t$, in order to keep the flame in the domain till the end $t_{III}$ of the simulations. Here, $S_t$ is the turbulent flame speed. Solely data obtained at $t_{II} \le t \le t_{III}$ are discussed in the following.

Two cases H and L characterized by High and Low, respectively, density ratios will be investigated. In case H, $\sigma = 7.53$, $S_L = 0.6$ m/s, $\delta_L = 0.217$ mm, $Da = 18$, $Ka = 0.21$, $S_t = 1.15$ m/s. In case L, $\sigma = 2.5$, $S_L = 0.416$ m/s, $\delta_L = 0.158$ mm, $Da = 17.3$, $Ka = 0.30$, and $S_t = 0.79$ m/s. Here, $\delta_L = (T_2 - T_1)/\max\{|\nabla T|\}$ is the laminar flame thickness, $Da = (L/u')/(\delta_L/S_L)$ and $Ka = (u'/S_L)^{3/2}(L/\delta_L)^{-1/2}$ are the Damköhler and Karlovitz numbers, respectively, evaluated at the leading edges of the mean flame brushes. The two flames are well associated with the flamelet combustion regime [23], e.g., various Bray-Moss-Libby (BML) expressions [44] hold, see figures 1-4 in Ref. [24].

## 3. Methods

### 3.1. Decomposition of velocity field

If velocity field $\mathbf{u}(\mathbf{x}, t)$ is decomposed into solenoidal and potential subfields $\mathbf{u}_s(\mathbf{x}, t)$ and $\mathbf{u}_p(\mathbf{x}, t)$, respectively, then,

$$\nabla \times \mathbf{u}_s = \nabla \times \mathbf{u}, \qquad \mathbf{u}_p = \nabla \varphi, \qquad \mathbf{u} = \mathbf{u}_s + \mathbf{u}_p. \tag{1}$$

However, such a decomposition is not unique, unless the function $\varphi(\mathbf{x}, t)$ is defined. To solve the problem, an extra constraint should be invoked and there are different methods for doing so. In the present work, two such methods, i.e. (i) widely-used "orthogonal" [45,46] and (ii) recently introduced "natural" [47,48] Helmholtz-Hodge decompositions, are used.

*3.1.1. Orthogonal decomposition*

The orthogonal decomposition invokes the following bulk constraint of orthogonality

$$\iiint_V \mathbf{u}_s \cdot \mathbf{u}_p d\mathbf{x} = 0 \tag{2}$$

of the subfields $\mathbf{u}_s(\mathbf{x}, t)$ and $\mathbf{u}_p(\mathbf{x}, t)$. This constraint results in the additivity of the bulk kinetic energies of the solenoidal and potential flow fields, i.e.

$$\iiint_V \mathbf{u} \cdot \mathbf{u} d\mathbf{x} = \iiint_V \mathbf{u}_s \cdot \mathbf{u}_s d\mathbf{x} + \iiint_V \mathbf{u}_p \cdot \mathbf{u}_p d\mathbf{x}, \tag{3}$$

but does not require that mean local correlation $\langle u_{s,i}(\mathbf{x}, t) u_{p,j}(\mathbf{x}, t) \rangle = 0$. Here, integration is performed over the computational domain $V$. Substitution of Eq. (1) into Eq. (2) yields

$$\iiint_V \mathbf{u}_s \cdot \nabla \varphi d\mathbf{x} = \iiint_V \nabla(\varphi \mathbf{u}_s) d\mathbf{x} - \iiint_V \varphi \nabla \cdot \mathbf{u}_s d\mathbf{x}$$

$$= \oiint_S \varphi \mathbf{u}_s \cdot \mathbf{n} dS - \iiint_V \varphi \nabla \cdot \mathbf{u}_s d\mathbf{x}, \tag{4}$$

where $S$ is the boundary of the domain $V$ and the unit vector $\mathbf{n}$ is normal to the boundary. If $\mathbf{u}_s \cdot \mathbf{n} = 0$ and $\nabla \cdot \mathbf{u}_s = 0$, then integrals in Eq. (4) vanish,

$$\Delta \varphi = \nabla \cdot \mathbf{u}_p = \nabla \cdot \mathbf{u} \tag{5}$$

in the entire domain $V$, and

$$\left. \frac{\partial \varphi}{\partial n} \right|_S = \mathbf{n} \cdot \nabla \varphi |_S = \mathbf{n} \cdot \mathbf{u}|_S \tag{6}$$

on its boundary. The Neumann problem given by Eqs. (5) and (6) has a unique solution for $\varphi(\mathbf{x}, t)$.

*3.1.2. Natural decomposition*

The natural decomposition [47,48] introduces an extra vector-field $\mathbf{w}(\mathbf{x}, t)$, which (i) coincides with the velocity field $\mathbf{u}(\mathbf{x}, t)$ in the domain $V$, i.e., $\mathbf{w}(\mathbf{x}, t) = \mathbf{u}(\mathbf{x}, t)$ for all $\mathbf{x} \in V$, but (ii) is extrapolated to the entire 3D space $\mathbb{R}^3$ such that $|\mathbf{w}(\mathbf{x}, t)| \to 0$ for $|\mathbf{x}| \to \infty$. Subsequently, the velocity field $\mathbf{w}(\mathbf{x}, t)$ is decomposed as follows

$$\mathbf{w} = \nabla\Gamma + \nabla \times \mathbf{S}, \qquad \mathbf{x} \in \mathbb{R}^3. \tag{7}$$

Consequently,

$$\Delta\Gamma = \nabla \cdot \mathbf{w}, \qquad \mathbf{x} \in \mathbb{R}^3, \tag{8}$$

$$\nabla \times \nabla \times \mathbf{S} = \nabla \times \mathbf{w}, \qquad \mathbf{x} \in \mathbb{R}^3. \tag{9}$$

Solutions to Poisson Eqs. (8) and (9) are unique

$$\Gamma(\mathbf{x}_0, t) = \iiint_{\mathbb{R}^3} G_\infty(\mathbf{x}, \mathbf{x}_0)\nabla \cdot \mathbf{w}(\mathbf{x}, t)d\mathbf{x}, \qquad \mathbf{x}_0, \mathbf{x} \in \mathbb{R}^3, \tag{10}$$

$$\mathbf{S}(\mathbf{x}_0, t) = -\iiint_{\mathbb{R}^3} G_\infty(\mathbf{x}, \mathbf{x}_0)\nabla \times \mathbf{w}(\mathbf{x}, t)d\mathbf{x}, \qquad \mathbf{x}_0, \mathbf{x} \in \mathbb{R}^3. \tag{11}$$

where $G_\infty(\mathbf{x}, \mathbf{x}_0) = -1/(4\pi|\mathbf{x}-\mathbf{x}_0|)$ is the free-space Green's function in $\mathbb{R}^3$. Finally, integration in Eqs. (10) and (11) is truncated outside the domain $V$ by interpreting the truncated integrals to be an external influence. Thus,

$$\Gamma^*(\mathbf{x}_0, t) = \iiint_V G_\infty(\mathbf{x}, \mathbf{x}_0)\nabla \cdot \mathbf{u}(\mathbf{x}, t)d\mathbf{x}, \qquad \mathbf{x}_0, \mathbf{x} \in V, \tag{12}$$

$$\mathbf{S}^*(\mathbf{x}_0, t) = -\iiint_V G_\infty(\mathbf{x}, \mathbf{x}_0)\nabla \times \mathbf{u}(\mathbf{x}, t)d\mathbf{x}, \qquad \mathbf{x}_0, \mathbf{x} \in V. \tag{13}$$

Obviously, Eq. (1) with $\mathbf{u}_s = \nabla \times \mathbf{S}^*$ and $\mathbf{u}_p = \nabla\Gamma^*$ holds by virtue of Eqs. (7)-(9).

*3.2. Conditioned structure functions*

Since the DNS data analyzed here were obtained from fully-developed, statistically 1D, planar flames that propagated from right to left along the $x$-direction, the computed flow fields are considered to be statistically homogeneous and isotropic in any transverse plane $x =$const. Accordingly, the following discussion will be restricted to SFs measured for two points $\mathbf{x}_A = \{x_{AB}, y_A, z_A\}$ and $\mathbf{x}_B = \{x_{AB}, y_A + r_y, z_A + r_z\}$ that belong to the same transverse plane $x = x_{AB}$, i.e., $\mathbf{x}_B - \mathbf{x}_A \equiv \mathbf{r} = \{0, r_y, r_z\}$. Then, the conditioned second-order SFs of the velocity field are defined as follows [12,13]

$$D_{ij}^{\alpha\beta}(x, r) \equiv \langle (u_{B,i} - u_{A,i})(u_{B,j} - u_{A,j})I_B^\alpha I_A^\beta\rangle / P_{\alpha\beta}, \tag{14}$$

where $\langle \cdot \rangle$ designates averaging over a transverse plane and time; $u_i$ is $i$-th component of the velocity vector $\mathbf{u} = \{u, v, w\}$; subscripts $i$ and $j$ refer to spatial coordinates; superscripts $\alpha$ and $\beta$ refer to the mixture state; the indicator function $I_A^u = I^u(\mathbf{x}_A, t)$ is equal to unity if unburned mixture is observed in point $\mathbf{x}_A$ at instant $t$ and vanishes otherwise; $I_A^b = 1$ if combustion products are observed in point $\mathbf{x}_A$ at instant $t$ and vanishes otherwise; $I_A^r = 1$ if $I_A^u = I_A^b = 0$ and vanishes otherwise; and $P_{\alpha\beta}(x, r) = \langle I_B^\alpha I_A^\beta \rangle$ are the probabilities that the mixture states $\alpha$ and $\beta$ are recorded in points $\mathbf{x}_B$ and $\mathbf{x}_A$, respectively, at the same instant. Depending on $\alpha$ and $\beta$, there are different conditioned SF tensors. The following discussion will be restricted to conditioned SFs $D_{ij}^{uu}$ (unburned mixture in both points). Henceforth, superscript $uu$ will be skipped for brevity.

When processing the DNS data, mean quantities $\bar{q}(x)$ are averaged over transverse $yz$-planes and over time (220 and 200 snapshots in cases H and L, respectively, stored during a time interval of $t_{III} - t_{II} \approx 1.5 L/u' \approx 10$ ms). Subsequently, $x$-dependencies are mapped to $\bar{c}$-dependencies using the spatial profiles $\bar{c}(x)$ of the Reynolds-averaged combustion progress variable. The probability $P_{uu}(x, r)$ and the unburned-mixture-conditioned SF are sampled from points characterized by $c(\mathbf{x}, t) < 0.05$, with Eq. (14) being separately applied to the potential or solenoidal velocity subfield to extract three pairs of conditioned SFs. More specifically, first, the transverse SFs $D_{xx,T}(x, r)$ for the axial solenoidal and potential velocities are obtained by averaging $(u_B - u_A)^2$ over time and two sets of points; (i) $\mathbf{x}_A = \{x, y, z\}$, $\mathbf{x}_B = \{x, y + r, z\}$ and (ii) $\mathbf{x}_A = \{x, y, z\}$, $\mathbf{x}_B = \{x, y, z + r\}$. To do so, both transverse coordinates are independently varied in intervals of $0 \leq y < \Lambda_y$ and $0 \leq z < \Lambda_z$, whereas the separation distance is varied in an interval of $0 \leq r < \Lambda_y/2 = \Lambda_z/2$. Second, the transverse SFs for the transverse velocities $D_{yz,T}(x, r) = 0.5(D_{yy,T} + D_{zz,T})$, where $D_{yy,T}$ and $D_{zz,T}$ are obtained by averaging (i) $(v_B - v_A)^2$ and (ii) $(w_B - w_A)^2$, respectively, over time and (i) $\mathbf{x}_A = \{x, y, z\}$, $\mathbf{x}_B = \{x, y, z + r\}$ and (ii) $\mathbf{x}_A = \{x, y, z\}$, $\mathbf{x}_B = \{x, y + r, z\}$, respectively. Third, the longitudinal SFs for the transverse velocities $D_{yz,L}(x, r) = 0.5(D_{yy,L} + D_{zz,L})$, where $D_{yy,L}$ and $D_{zz,L}$ are obtained by averaging (i) $(v_B - v_A)^2$ and (ii) $(w_B - w_A)^2$, respectively, over time and (i) $\mathbf{x}_A = \{x, y, z\}$, $\mathbf{x}_B = \{x, y + r, z\}$ and (ii) $\mathbf{x}_A = \{x, y, z\}$, $\mathbf{x}_B = \{x, y, z + r\}$, respectively.

## 4. Results and Discussion

Figure 1a shows that, upstream of the mean flame brush (0.5 mm $< x <$ 1 mm), the mean solenoidal kinetic energy $\overline{(\mathbf{u}_s - \overline{\mathbf{u}_s})^2}/2$ (curves 1 and 4) is significantly higher than the mean potential kinetic energy $\overline{(\mathbf{u}_p - \overline{\mathbf{u}_p})^2}/2$ (curves 2 and 5) and is close to the mean kinetic energy $\overline{(\mathbf{u} - \overline{\mathbf{u}})^2}/2$ (curve 3). Both $\overline{(\mathbf{u}_s - \overline{\mathbf{u}_s})^2}/2$ and $\overline{(\mathbf{u} - \overline{\mathbf{u}})^2}/2$ decay in the $x$-direction in the unburned mixture ($x <$ 1 mm), but begin increasing within the mean flame brush. On the contrary, the potential $\overline{(\mathbf{u}_p - \overline{\mathbf{u}_p})^2}/2$ begins increasing already in the unburned mixture upstream of the mean flame brush ($x \approx 0.5$ mm) and exceeds the solenoidal $\overline{(\mathbf{u}_s - \overline{\mathbf{u}_s})^2}/2$ in the largest part of the mean flame brush (1.3 mm $< x <$ 3 mm or $0.1 < \bar{c} < 0.86$). In case L characterized by a low density ratio, both $\overline{(\mathbf{u}_s - \overline{\mathbf{u}_s})^2}/2$ and $\overline{(\mathbf{u} - \overline{\mathbf{u}})^2}/2$ decrease with $x$, see Fig. 1b, but there is a local peak of the kinetic energy of the potential motion within the mean flame brush so that $\overline{(\mathbf{u}_p - \overline{\mathbf{u}_p})^2}$ and $\overline{(\mathbf{u}_s - \overline{\mathbf{u}_s})^2}$ are comparable there. These results show generation of potential velocity fluctuations not only within the mean flame brush, but also in the unburned mixture upstream of the mean flame brush (at least in case H), with the effect magnitude being significantly increased by the density ratio. Note that generation of potential velocity perturbations is well known in the theory of the hydrodynamic instability of a laminar premixed flame [49], but solenoidal velocity vanishes upstream of the flame in that case. On the contrary, solenoidal velocity fluctuations dominate upstream of mean flame brushes in the present study.

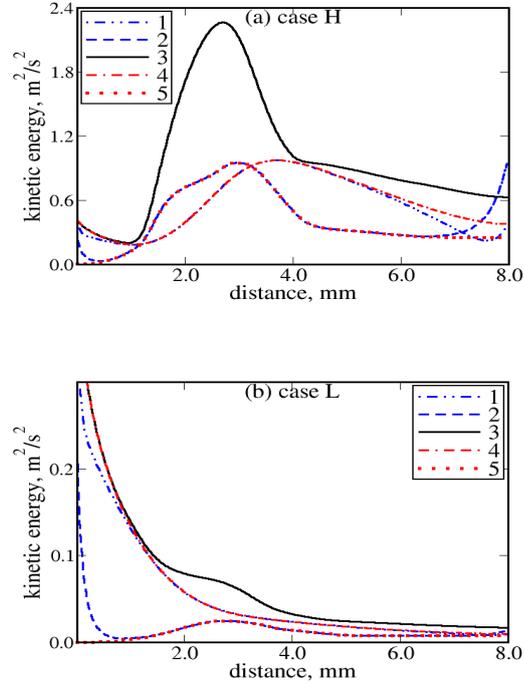

**Fig. 1.** Contributions of solenoidal (curves 1 and 4) and potential (curves 2 and 5) velocity fields to the mean kinetic energy (curve 3) of velocity fluctuations. Results obtained using orthogonal (natural) decomposition are shown in curves 1 and 2 (4 and 5, respectively).

To further explore the influence of combustion-induced thermal expansion on turbulence in unburned mixture upstream of the flame, let us consider behavior of SFs conditioned to the unburned mixture. Since comparison of curves 1 and 2 in Fig. 1 with curves 4 and 5, respectively, indicates that the two adopted Helmholtz-Hodge decompositions yield close results in the largest part of the computational domain with the exception of regions near the inlet and outlet boundaries, we will restrict ourselves to reporting data obtained using the natural decomposition. It is worth stressing that all trends discussed in the following are also observed when using the orthogonal decomposition.

Figure 2 shows that longitudinal SFs $D_{yz,L}$ for solenoidal transverse velocities, normalized using conditioned solenoidal transverse rms velocity $(\overline{v'^2_{s,u}} + \overline{w'^2_{s,u}})/2$, decrease slowly with $x$, as expected for spatially decaying turbulence. Similar results were obtained when normalizing these SFs with the total $(\overline{v'^2_u} + \overline{w'^2_u})/2$. On the contrary, for the potential velocity field, such SFs normalized with $(\overline{v'^2_u} + \overline{w'^2_u})/2$ increase significantly with $x$ in both cases, see Figs. 3a and 3b, with the effect being observed well upstream of the mean flame brush, see curves 1-3. If the potential $D_{yz,L}$ is normalized using the potential $(\overline{v'^2_{p,u}} + \overline{v'^2_{p,u}})/2$, the evolution of such SFs is weakly pronounced at $\bar{c} \leq 0.1$, see curves 1-4 in Figs. 3c and 3d. Thus, Fig. 3 implies that potential perturbations of the transverse velocity field in unburned mixture are mainly controlled by an increase in the magnitude $\left(\overline{v'^2_{p,u}} + \overline{v'^2_{p,u}}\right)/2$ of such perturbations, whereas changes in the spatial scales of the perturbations are weakly pronounced. However, it is worth noting that, at large $r$, the solenoidal $2D_{yz,L}/\left(\overline{v'^2_{s,u}} + \overline{v'^2_{s,u}}\right) \to 2$, see Fig. 2, but the potential $2D_{yz,L}/\left(\overline{v'^2_{p,u}} + \overline{v'^2_{p,u}}\right) \to 4$, see Figs. 3c and 3d, thus, indicating that, contrary to a typical turbulent flow, where $D_{ij}(r \to \infty) \to 2\overline{u'_i u'_j}$ [50], the discussed potential velocity perturbations can be negatively correlated at large distances.

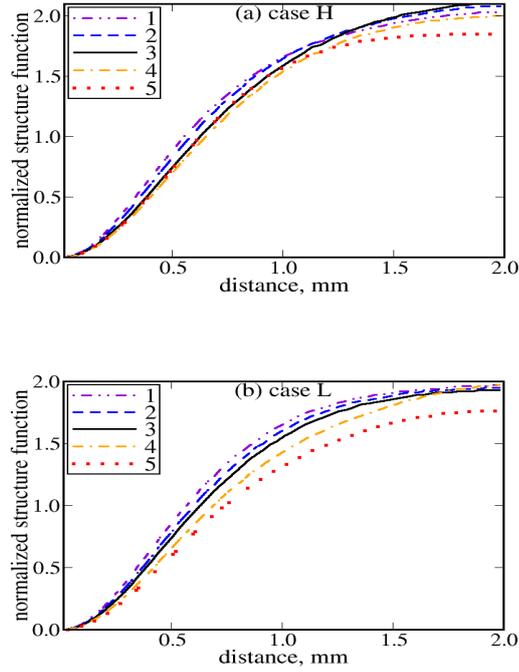

**Fig. 2.** Longitudinal SFs $D_{yz,L}$ for solenoidal transverse velocities, conditioned to unburned mixture and normalized using conditioned solenoidal transverse rms velocity $\left(\overline{v'^2_{s,u}} + \overline{w'^2_{s,u}}\right)/2$. (*a*) case H, (*b*) case L. $1 - \bar{c} = 0, x = 0.50$ mm; $2 - \bar{c} = 0, x = 0.75$ mm; $3 - \bar{c} = 0.01, x = 1.0$ mm; $4 - \bar{c} = 0.1, x = 1.3$ or 1.4 mm in case H or L, respectively; $5 - \bar{c} = 0.25, x = 1.5$ or 1.8 mm in case H or L, respectively.

Results computed for transverse SFs $D_{yz,T}$ and $D_{xx,T}$, see Figs. 4-5 and 6-7, respectively, are qualitatively similar to the already discussed results for the longitudinal $D_{yz,L}$, but there are some differences. First, the magnitudes of the potential $2D_{yz,T}/\left(\overline{v'^2_u} + \overline{w'^2_u}\right)$ or $D_{xx,T}/\overline{u'^2_u}$ are less than the magnitudes of the potential $2D_{yz,L}/\left(\overline{v'^2_u} + \overline{w'^2_u}\right)$, cf. Fig. 4 or 6a and Figs. 3a-3b, respectively, with the effect being strongly pronounced in case L, cf. Fig. 3b and 4b.

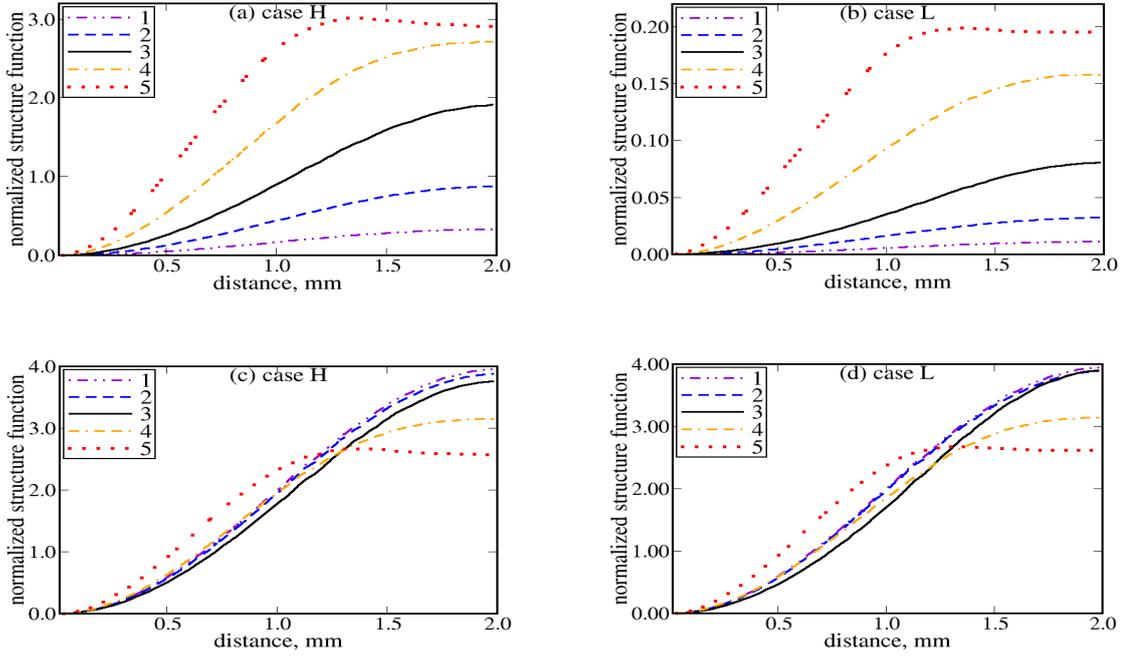

**Fig. 3.** Longitudinal SFs $D_{yz,L}$ for potential transverse velocities, conditioned to unburned mixture and normalized using conditioned total (*a* and *b*) or potential (*c* and *d*) transverse rms velocity $\left(\overline{v'^2_u} + \overline{w'^2_u}\right)/2$ or $\left(\overline{v'^2_{p,u}} + \overline{v'^2_{p,u}}\right)/2$, respectively. (*a*) and (*c*) case H, (*b*) and (*d*) case L. Legends are explained in caption to Fig. 2.

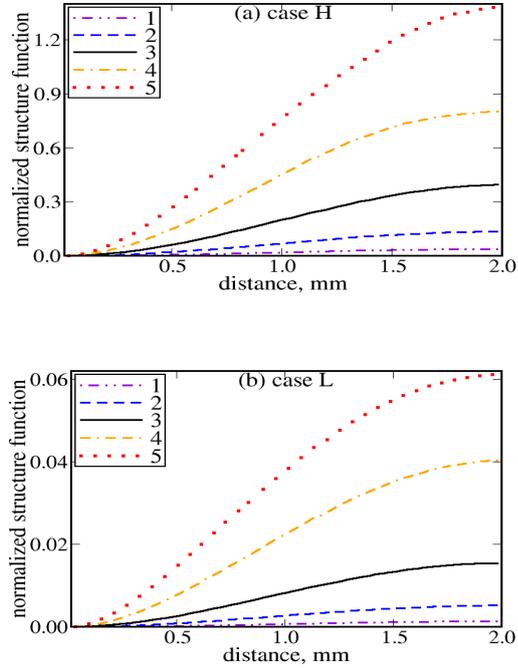

**Fig. 4.** Transverse SFs $D_{yz,T}$ for potential transverse velocities, conditioned to unburned mixture and normalized using conditioned transverse rms velocity $\left(\overline{v'^2_u} + \overline{w'^2_u}\right)/2$. (*a*) case H, (*b*) case L. Legends are explained in caption to Fig. 2.

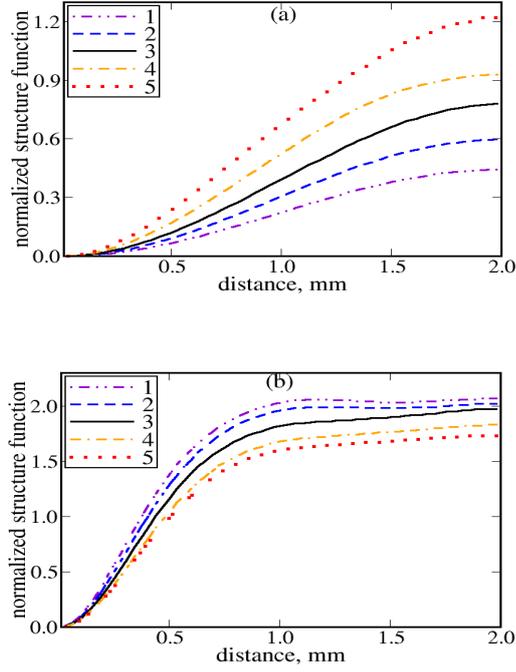

**Fig. 5.** Transverse SFs $D_{yz,T}$ for (*a*) potential and (*b*) solenoidal transverse velocities, conditioned to unburned mixture and normalized using conditioned (*a*) potential or (*b*) solenoidal transverse rms velocities $\left(\overline{v'^2_{p,u}} + \overline{w'^2_{p,u}}\right)/2$ or $\left(\overline{v'^2_{s,u}} + \overline{w'^2_{s,u}}\right)/2$, respectively. Case H. Legends are explained in caption to Fig. 2.

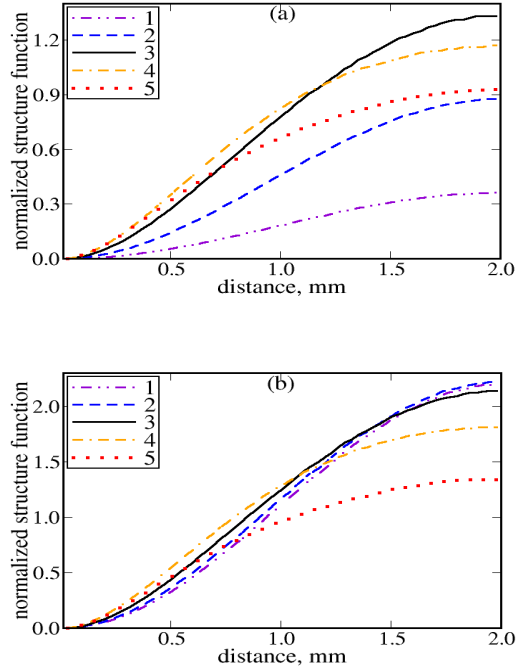

**Fig. 6.** Transverse SFs $D_{xx,T}$ for potential axial velocity, conditioned to unburned mixture and normalized using conditioned (*a*) total or (*b*) potential axial rms velocity $\overline{u'^2_u}$ or $\overline{u'^2_{p,u}}$, respectively. Case H. Legends are explained in caption to Fig. 2.

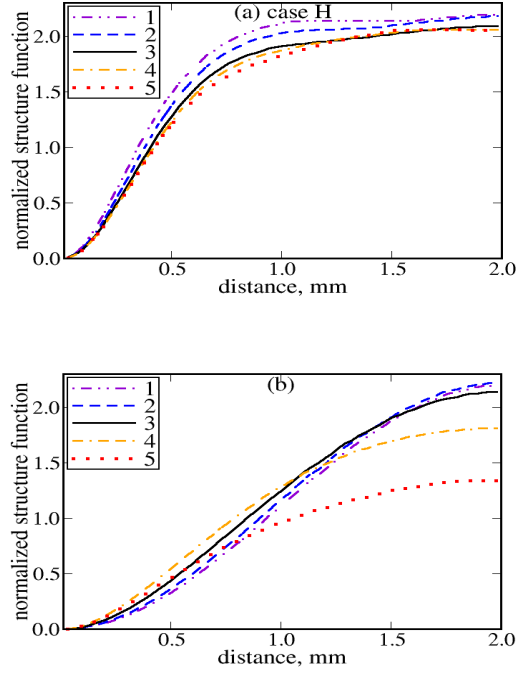

**Fig. 7.** Transverse SFs $D_{xx,T}$ for solenoidal axial velocity, conditioned to unburned mixture and normalized using conditioned solenoidal axial rms velocity $\overline{u'^2_{s,u}}$ (a) case H, (b) case L. Legends are explained in caption to Fig. 2.

Second, the potential $D_{yz,T}$ normalized using the potential $\overline{(v'^2_{p,u} + w'^2_{p,u})}/2$ increases with $x$, but remains smaller than 2, see Fig. 5a, whereas the potential $D_{xx,T}/\overline{u'^2_{p,u}}$ evolves slowly with $x$ if $\bar{c} \leq 0.1$, see curves 1-4 in Fig. 6b, similarly to $2D_{yz,L}/\overline{(v'^2_{p,u} + w'^2_{p,u})}$ see Fig. 3c. Moreover, the magnitudes of the normalized potential transverse SFs $2D_{yz,T}/\overline{(v'^2_{p,u} + w'^2_{p,u})}$ and $D_{xx,T}/\overline{u'^2_{p,u}}$ are significantly less than the magnitude of the normalized potential longitudinal SF $2D_{yz,L}/\overline{(v'^2_{p,u} + w'^2_{p,u})}$, cf. Figs. 5a and 6b with Fig. 3c. Thus, the spatial structure of the potential velocity fluctuations generated in unburned mixture upstream of the studied flames is highly anisotropic.

Third, a decrease in the solenoidal SF $2D_{yz,T}/\overline{(v'^2_{s,u} + w'^2_{s,u})}$ or $D_{xx,T}/\overline{u'^2_{s,u}}$ with $x$, see Fig. 5b or 7a, respectively, is better pronounced when compared to the evolution of $2D_{yz,L}/\overline{(v'^2_{s,u} + w'^2_{s,u})}$ with $x$, see Fig. 2a.

## 5. Conclusions

Helmholtz-Hodge decomposition and conditioned second-order structure functions of turbulent velocity field were jointly (for the first time to the best of the present authors' knowledge) applied to processing DNS data obtained earlier from two weakly turbulent premixed flames characterized by different density ratios. Computed results show that combustion-induced thermal expansion can significantly change turbulent flow of unburned mixture upstream of a premixed flame by generating anisotropic potential velocity fluctuations whose spatial structure differ substantially from spatial structure of the incoming turbulence. On the contrary, the influence of the thermal expansion on the solenoidal-velocity-field structure functions in the unburned mixture is of minor importance under conditions of the present study. Assessment of the effect of an increase in $u'/S_L$ or Karlovitz number on the magnitude of the revealed effects appears to be of great fundamental interest for future studies.

.